\documentclass[useAMS,usenatbib,]{mn2e}

\include{epsf}
\usepackage{graphicx}
\usepackage{times}
\newcommand{\hi}{H\,{\sc i}}
\newcommand{\oii}{O\,{\sc ii}}
\newcommand{\ha}{H{$\alpha$}}

\newcommand{\msol}{\mbox{${\rm M}_\odot$}}
\newcommand{\hubble}{\mbox{$\rm km\, s^{-1}\, Mpc^{-1}$}}
\newcommand{\kms}{\mbox{$\rm km\, s^{-1}$}}
\newcommand{\jykms}{\mbox{$\rm Jy\, km\, s^{-1}$}}
\newcommand{\mhi}{\mbox{$M_{\rm HI}$}}
\newcommand{\sint}{\mbox{$S_{\rm int}$}}
\newcommand{\speak}{\mbox{$S_{\rm p}$}}

\title[The cold gas content of post-starburst galaxies]{The cold gas content of post-starburst galaxies}
\author[M. A. Zwaan et al.]{
Martin. A. Zwaan,$^{1}$\thanks{E-mail: mzwaan@eso.org}
Harald Kuntschner,$^{1}$
Michael B. Pracy,$^{2}$ and
Warrick J. Couch$^{3}$\\
\\
$^{1}$European Southern Observatory, Karl-Schwarzschild-Str. 2, 85748 Garching
     b. M{\"u}nchen, Germany\\
$^{2}$Sydney Institute for Astronomy, School of Physics, University of Sydney, NSW 2006, Australia\\     
$^{3}$Center for Astrophysics \& Supercomputing, Swinburne University of Technology, P.O. Box 218, Hawthorn, Vic, Australia\\
}

\begin{document}

\date{Accepted ...
        Received  ...}

\pagerange{\pageref{firstpage}--\pageref{lastpage}} \pubyear{2011}

\maketitle

\label{firstpage}

\begin{abstract}
Post-starburst galaxies, or E+A galaxies, are characterized by optical spectra showing strong Balmer absorption lines, indicating a young stellar population, and little or no emission lines, implying no active star formation. These galaxies are interpreted as a transitional population between star-forming, disk-dominated galaxies and spheroidal quiescent, non-star forming galaxies. Here, we present single dish \hi\ 21-cm emission line measurements of a sample of eleven of these galaxies at redshifts $z<0.05$. We detect \hi\ emission in six of the E+A galaxies. In combination with earlier studies, the total number of E+A galaxies with measured cold gas components is now eleven. Roughly half of the E+As studied so far have detectable \hi. The gas fractions of these galaxies, measured with respect to their stellar mass, are between 1 and 10 percent and are at the high end of the gas fractions measured in gas-bearing early type galaxies and typically lower than seen in late-type galaxies with comparable stellar masses. This finding is consistent with the idea that E+As are currently evolving from the blue cloud to the red sequence. However, the question of why the star formation has ceased in these galaxies while a significant gas reservoir is still present can only be answered by higher spatial resolution observations of the cold gas. 

\end{abstract}


\section{Introduction}
Recent large scale redshift surveys have established that the local galaxy distribution is bimodal, showing a red sequence consisting of spheroidal quiescent, non-star forming, gas-poor galaxies and a blue cloud populated by morphologically disk-dominated, star-forming, gas-rich galaxies \citep[e.g.][and references therein]{Blanton2009a}. While the bimodality has been in place since redshifts of at least $z=2$ \citep{Williams2009a}, it has been shown that the blue cloud has evolved little since that time, while the red sequence has increased in mass by at least a factor two \citep{Bell2004a,Faber2007a}. Over time, blue galaxies that quench their star formation, move off the blue cloud and evolve into red sequence galaxies. Galaxies that are in this transition phase are located in the so-called "green valley" of the colour-magnitude diagram. The fact that the blue and red populations are very pronounced in the colour-magnitude diagram implies that the time that  galaxies spend in the green valley must be relatively short  ($\sim$1~Gyr), or else the bimodality would wash out \citep{Tinker2010a}.

E+A galaxies (or post-starburst galaxies) are a crucial part of this picture. Observational evidence indicates that the E+A phenomenon may, in part, mark the transitional phase between star-forming disk galaxies and quiescent spheroidal systems \citep[e.g.,][]{Caldwell1996a,Zabludoff1996a, Norton2001a, Pracy2009a}.  Their optical spectra, characterized by strong Balmer absorption lines (implying a relatively young A-type stellar population) and an absence of [\oii] and \ha\ emission lines, represent galaxies undergoing rapid evolution in their star formation properties. The A-stars must have been produced by a powerful recent episode of star formation, which has been suddenly truncated in the past $\sim$1 Gyr \citep{Couch1987a}. 

At intermediate redshifts, many E+A galaxies are found in galaxy clusters: at redshifts $z\sim 0.4$ typically up to 20\% of the cluster population are found to consist of E+As \citep{Dressler1983a,Couch1998a,Dressler1999a,Poggianti1999a}.
However most E+A galaxies at low redshift are located in the field and represent a very low fraction of the overall galaxy population (0.2\%; Zabludoff et al. 1996). The rarity of E+A galaxies at low redshift means that samples can only be constructed from large galaxy redshift surveys. As a result, very few local ($v < 5000$~km/s) examples of field E+As are known. 

For these local E+A galaxies, galaxy mergers (followed by a starburst) have been implicated as a plausible formation mechanism, evidenced by numerous examples of tidal signatures and disturbed morphologies \citep{Zabludoff1996a, Blake2004a,Yang2008a, Pracy2009a}. Spatially resolved kinematics and stellar population gradient studies point to mergers being the dominant mechanism \citep{Pracy2012a,Pracy2013a}. 
In particular,  negative Balmer gradients in the centres of E+As indicate young central stellar  merger-induced star bursts. Consistent with the overall population of early-type galaxies \citep{Emsellem2011a}, about 85 per cent of the E+A galaxies are fast rotators showing rather regular stellar rotation with approximately aligned photometric and kinematic axes \citep{Pracy2013a}.

The presence or lack of substantial reservoirs of cold gas places powerful constraints on both the formation and future evolution of E+A galaxies. If the truncation of star formation is caused by the complete removal of the gas supply (via stripping or complete exhaustion of the supply during a starburst), then the E+A phase represents the last star-formation episode in a galaxy. If, however, the truncation of star formation is a temporary effect caused by the starburst but the gas reservoirs are left in place, then the E+A phase may be transient with another bout of star formation to come. 

The study of the cold gas content of E+A galaxies should follow a staged approach. First, it needs to be established what fraction of E+A galaxies has a detectable amount of cold-gas. This question can best be addressed by either single dish or low spatial resolution 21-cm synthesis observations of low redshift E+A galaxies. At present only a few of these searches for cold gas have been conducted. In only five E+A galaxies the neutral hydrogen HI 21-cm line has been measured \citep{Chang2001a, Buyle2006a}. This limited number of detections is mostly due to the fact that the majority of previously studied E+A galaxies are at large distances, and therefore required unrealistically long integration times on present-day radio telescopes. In the present paper, we present a new survey for \hi\ in E+A galaxies using the NRAO Green Bank Telescope (GBT) which resulted in six new detections of \hi\ in E+A galaxies. The second stage of the study of cold gas in these galaxies should be higher spatial resolution 3-dimensional mapping of the gas. Higher resolution 21-cm mapping can establish, for example, the presence of tidal features indicating galaxy interactions. Furthermore, observations of the molecular gas phase can be used to investigate whether the molecular gas densities are sufficiently high to start new episodes of star formation. To this end, we have initiated a program of mapping E+A galaxies in the 21-cm line with the Karl G. Jansky Very Large Array (JVLA) and in the CO(1-0) line with Atacama Large Millimeter/submillimeter Array (ALMA). These results will be presented in a forthcoming paper.

This paper is organized as follows. In the next section we describe the selection of our sample of nearby E+A target galaxies, followed by a description of the observations and data reduction in Section 3. In Section 4 we present the results and compare the gas-richness of E+A galaxies with that of blue cloud and red sequence galaxies in the local Universe. Finally, conclusions and future work are discussed in Section 5.
For distance dependent quantities we use $H_0=70~\hubble$ 
\citep{Jarosik2011a}.

\section{Sample selection}
Our target galaxy sample is taken from two parent samples of E+A galaxies. The first sample is selected from the 2dF Galaxy Redshift Survey and a detailed study of their morphologies, environments, clustering properties and luminosity function has been presented in \citet{Blake2004a}. Essentially, the selection criterion for inclusion in this sample was either an H$\delta$ equivalent width (EW) of $>$ 5.5\AA\ in absorption, or a mean equivalent width of the H$\beta$, H$\gamma$, H$\delta$ lines (H$\beta\gamma\delta$) of $>$ 5.5\AA\ in absorption, in combination with no detectable [O II] 3727\AA\ emission. 
The second parent sample is that presented by \citet{Goto2007a}, carefully selected from half a million spectra from the Sloan Digital Sky Survey Data Release 5. The selection criteria for the Goto sample were 
EW(H$\delta$) $>5.0$ \AA\ in absorption together with low [OII] 3727\AA\ and H$\alpha$ emission: 
EW(H$\alpha$) $>-3.0$ \AA\ and EW([O II]) $>-2.5$ \AA.

\begin{table*}
\centering
\begin{minipage}{116mm}
\caption{The sample of 11 E+A galaxies observed with the GBT}
\begin{tabular}{l c c c c c }
\hline
 galaxy name &  source & $V^{\rm opt}_{\rm hel}$  &  $M_B$ & $M_K$ & environment \\
   &    &   (\kms) & (mag) & (mag) \\
   (1) & (2) & (3) & (4) & (5) & (6) \\  
\hline\hline
CGCG 017-079  			& 2dF &   4456 		&   	$-$18.4  &    $-$21.0 &  Poor cluster/group$^1$	\\
2dFGRS S259Z083 			& 2dF &   8497  	&	$-$20.0  &    $-$24.1 &  Field	 	\\
2dFGRS S833Z022  			& 2dF &   2193   	&  	$-$16.1  &        ... 	 &  Field		\\
ESO405-G015      			& 2dF &  9914    	&   	$-$20.4  &         ...	  &  Cluster$^2$		\\
ESO471-G032       			& 2dF &  8710    	&  	$-$19.4  &    $-$24.4 &  Cluster$^2$ 	\\
ESO534-G001       			& 2dF &  1407    	&   	$-$15.8  &         ... 	 &  Field		\\
IC0976      				& SDSS &  1525    	&   	$-$17.7  &    $-$21.7 &  Group$^3$	 	\\
MRK 0779  				& SDSS &  12249 	&   	$-$20.5  &    $-$24.4 &  Field	 	\\
MRK 0239   				& SDSS &  11399  	&  	$-$20.6  &    $-$24.3 &  Poor cluster/group$^1$	 	\\
SDSS J145455.45+453126.6 	& SDSS &  10979  	&  	$-$19.9  &    $-$23.9 &  Field	 	\\
SDSS J162702.64+432833.9  	& SDSS &  13865 	&  	$-$20.8  &    $-$24.9 &  Cluster/infall region$^4$	 	\\
\hline
\end{tabular}
\label{sample.tab}
\\{\small {\em Notes:} 
column (1): galaxy name;
column (2): parent sample from which the E+A galaxies are selected: '2dF' indicates the 2dF Galaxy Redshift Survey sample \citep{Blake2004a} and 'SDSS' indicates the Sloan Digital Sky Survey sample \citep{Goto2007a}  ;
column (3): heliocentric recessional velocity from NED;
column (4): absolute $B$-band magnitude;
column (5): absolute $K$-band magnitude taken from the 2MASS Point Source Catalogue where available \citep{Skrutskie2006a};
column (6): environment, indicating whether the galaxy is a cluster or group member, or in the field. References:
1: \citet{White1999a},
2: \citet{Lauberts1982a},
3: \citet{Garcia1993a},
4: \citet{Rines2002a}.
}
\end{minipage}
\end{table*}%

From this list we selected the 11 most promising galaxies to be observed in the \hi\ 21-cm line with the GBT. The selection was done in the following way: We aimed for a minimal detectable \hi\ mass equal to 10\% of that expected for spiral galaxies of similar luminosities. Using this detection limit, all E+A galaxies previously detected in the \hi\ 21-cm line should be detectable. For the relation between \hi\ mass and $B$-band luminosity for spiral galaxies we adopt from \citet{Zwaan2000c}: $\log \mhi = -0.25 M_B +4.5$. Using this relation and assuming the 10\% gas fraction, we derived an estimated \hi\ gas mass  for each E+A galaxy in the parent samples. Next we converted this \hi\ mass into an integrated 21-cm line flux, using the Hubble distances to the galaxies and sorted the list on descending expected integrated flux. We then selected the top 11 galaxies with declinations observable with the GBT.  All of our targets have redshifts $z<0.05$. The $B$-band absolute magnitudes range from $-15.8$ to $-20.9$. Table~\ref{sample.tab} summarizes the properties of our sample.

\begin{figure*}
\begin{center}
\includegraphics[width=16cm,trim=0.cm 7.0cm 0.cm 0.cm]{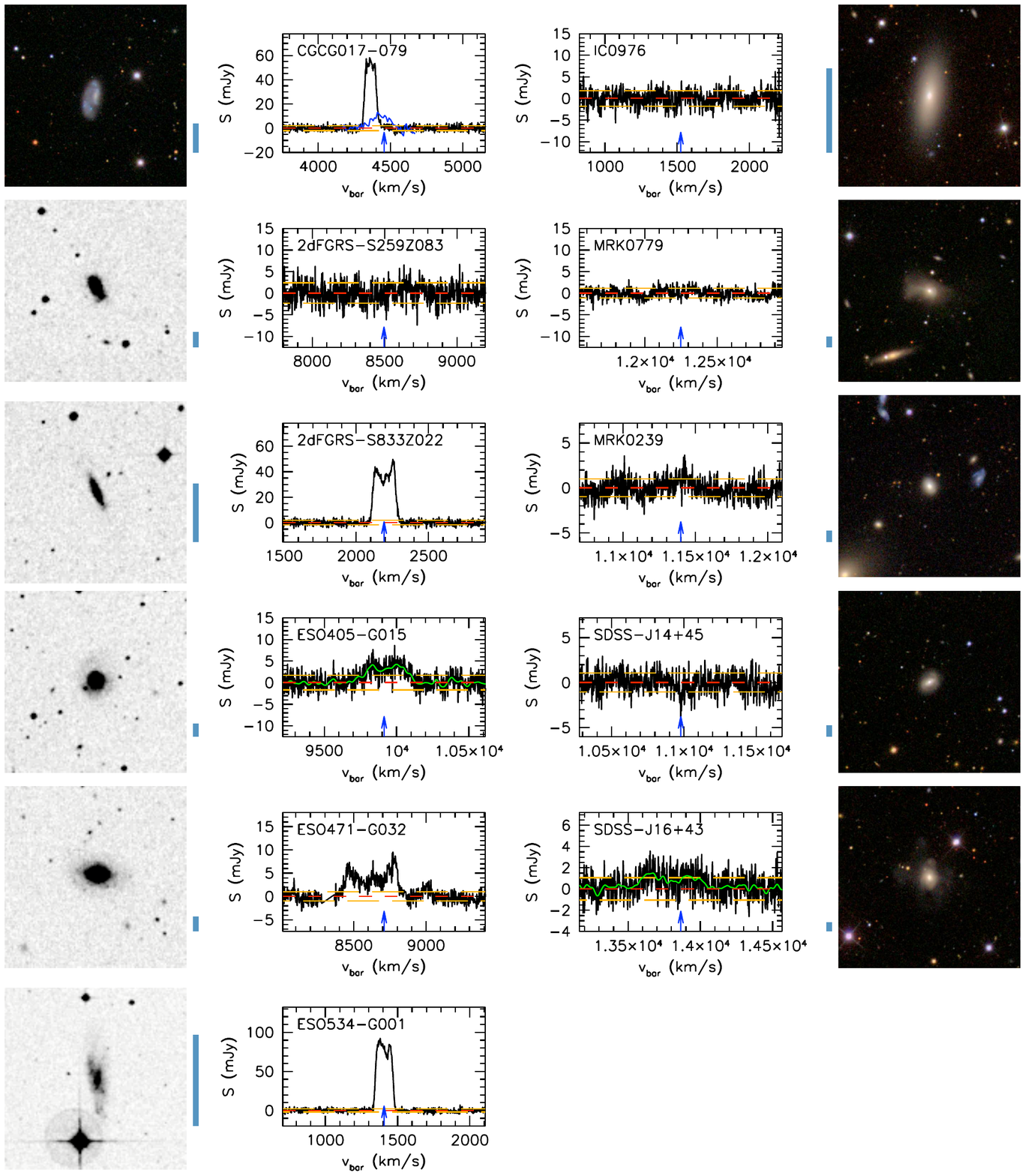}
\caption{{\em Spectra\/}: GBT \hi\ 21-cm spectra of the E+A galaxies observed. The arrows at the bottom of each spectrum show the optical redshifts of the galaxies. The thin green lines in the spectra of  ESO405-G015 and SDSS~J162702.64+432833.9 show spectra smoothed with a Gaussian kernel with FWHM=30~\kms. The dashed orange lines indicate the $1\sigma$ and $-1\sigma$ rms noise level. 
For CGCG~017-079 the GBT spectrum is confused by the \hi\ signal of a nearby galaxy. The blue line indicates the spectrum of CGCG~017-079 from higher spatial resolution JVLA data that we have obtained. {\em Images\/}:  $3.4' \times 3.4'$ images of the galaxies in our GBT sample. Greyscale images are from the Digitized Sky Survey, colour images are from the SDSS. The length of the blue scale bars next to each of the images corresponds to 10 kpc at the distance of the galaxies.
 \label{spectra.fig}}
\end{center}
\end{figure*}

\section{Observations and data reduction}
Observations were carried out with the NRAO Green Bank Telescope (GBT)  during September 2009. For the back-end we used the spectral processor employing a total bandwidth of 10 MHz, two polarizations, 1024 channels and a 5s integration time. This setup results in a spectral resolution of approximately $2.1\, \rm km\, s^{-1}$ over a total velocity range of $2100\, \rm km\, s^{-1}$, which is sufficiently wide to include all \hi\ emission of the target galaxies and allows for reliable baseline fitting. The receiver used for this program is the L-band receiver with $T_{\rm sys}=20$~K. The observations were taken in standard on-off mode, with 5 min on and 5 min off scans. On-source integration times ranged from 0.8 to 2.8 hours, depending on the redshift of the targets. 

Data reduction was carried out with the {\sc gbtidl} interactive data reduction package. This package allows for a flexible editing of the data in order to remove data sections with radio frequency interference (RFI) or obvious correlator problems. All data were scrutinized by eye and carefully edited. Calibration was done by subtracting the off-scans and by observing bright continuum sources with known flux density, all using standard procedures. Data taken for each object were then median filtered to obtain a single spectrum per object. 
Baselines were removed by making polynomial fits with orders between 2 and 5 to the RFI-free channels, and excluding 500~\kms\  around the optical redshifts of the galaxies (600~\kms\ in the case of SDSS J16+43). Residual structure can be seen in some of the spectra, but higher order polynomial fits would result in unreliable interpolations of the baselines across the galaxy spectra. We carefully checked that there was no \hi\ emission beyond the 500~\kms\ or 600~\kms\ exclusion regions around the optical redshifts.
The final r.m.s. noise levels are between $\sim$1 and $\sim$2.4 mJy per channel.
The final spectra are displayed in Fig.~\ref{spectra.fig}.

The spectra for the two systems that showed very low signal to noise detections were smoothed to $\approx 30$~\kms . These smoothed spectra are indicated by green lines in Fig.~\ref{spectra.fig}. Integrated \hi\ line fluxes were calculated by simply summing the fluxes in all channels within $\sim 50$~\kms\ of where the profiles reach  the baseline. If the flagging of RFI caused small gaps in the spectra, a linear interpolation was made to estimate the line flux in the flagged channels. Velocity widths at 20 and 50 per cent of the peak flux were determined from the spectra at full resolution, except for the two low signal-to-noise detections, in which case the smoothed spectra were used. The recessional velocities were measured from the \hi\ spectra by taking the mean of the velocities at which the flux reaches  50 per cent of the peak flux. The arrows at the bottom of each spectrum in Fig.~\ref{spectra.fig} show the optical redshifts of the galaxies.

\section{Results}
\label{results.sec}

\begin{table*}
\centering
\caption{Summary of results of the GBT observations of our sample of 11 E+A galaxies}
\begin{tabular}{l c c c c c c c c c c}
\hline
 galaxy name & $T_{\rm int}$ &  \sint\  &  \speak\ & \mhi\ &  $W_{20} $ &   $ W_{50} $ &  $V^{\rm HI}_{\rm hel}$  &  ch.sep. &  r.m.s. &   $M_{\rm HI, min}$$^2$ \\
   &      (min) &      (\jykms) & (mJy)  &   ($10^8$ \msol)  &   (\kms) &   (\kms)  &  (\kms)  &  (\kms) &  (mJy)  &    ($10^8$ \msol) \\
   (1) & (2) & (3) & (4) & (5) & (6) & (7) & (8) & (9) & (10) & (11)\\  
\hline\hline
CGCG 017-079$^1$ &  50 & 0.44 &  12.1 &  4.1 &    188 &     148 &   4437 &  2.12 &  2.05 &   2.34 \\ 
 2dFGRS-S259Z083 &   60 & ... &   ... &   ... &      ... &      ... &    ...   &  2.18 &  2.37 &  10.01 \\ 
 2dFGRS-S833Z022 &   60 & 6.35 &  42.3 &  14.7 &    184 &    167 &   2194 &  2.09 &  1.82 &   0.50 \\ 
 ESO405-G015 &  70 &  1.16 &   4.25 &  55.2 &    400 &    301 &   9938 &  2.20 &  1.73 &   9.95 \\ 
 ESO471-G032 &  145 &  1.66 &   6.96 &  59.7 &    395 &    384 &   8637 &  2.18 &  1.00 &   4.43 \\ 
 ESO534-G001 & 40 & 10.1 &  85.3 &   9.63 &    137 &    123 &   1407 &  2.08 &  2.01 &   0.23 \\ 
     IC0976 &  45 &  ... &   ... &   ... &      ... &      ... &   ...   &  2.08 &  1.78 &   0.24 \\ 
 MRK 0779 &  125 &  ... &   ... &   ... &      ... &      ... &    ...   &  2.23 &  1.14 &  10.13 \\ 
 MRK 0239 &   170 & ... &   ... &   ... &      ... &      ... &    ...   &  2.22 &  0.98 &   7.49 \\ 
 SDSS J145455.45+453126.6 & 140 &   ... &   ... &   ... &      ... &      ... &   ...    &  2.22 &  1.05 &   7.43 \\ 
 SDSS J162702.64+432833.9  &  105 &  0.48 &   1.51 &  44.0 &    496 &    419 &  13796 &  2.26 &  1.06 &  12.11 \\ 
\hline
\end{tabular}
\label{obstable.tab}
\\{\small {\em Notes:} 
column (1): galaxy name;
column (2): on-source integration time;
column (3): measured integrated \hi\ 21-cm flux;
column (4): measured peak \hi\ 21-cm flux;
column (5): derived \hi\ mass, assuming $H_0=70~\hubble$;
column (6): width of 21-cm profiles measured at 20 per cent of peak intensity;
column (7): width of 21-cm profiles measured at 50 per cent of peak intensity;
column (8): measured heliocentric velocity defined as the mean of the velocities where the profile reaches 50 per cent of peak flux ;
column (9): channel separation;
column (10): r.m.s. noise level per channel measured in the spectrum;
column (11): minimal detectable \hi\ mass (see text);
$^1$: Measurements of this galaxy are derived from follow-up JVLA observations that will be presented in a forthcoming paper. The GBT \hi\ signal is confused by the nearby galaxy NGC~5327.
$^2$: Measured from spectra smoothed to 30 \kms.
}
\end{table*}%

Table~\ref{obstable.tab} summarizes the observational results of our GBT observations. We detected \hi\ emission in 6 galaxies out of the 11 galaxies that were observed. The peak fluxes of our detections range from 1.5 to 85 mJy and integrated line fluxes range from 0.5 to 10 \jykms. Total \hi\ masses are calculated using the standard equation $\mhi=2.36\times 10^5 \sint D_{L}^2$, where $D_L$ is the luminosity distance measured in Mpc and \sint\ is the integrated \hi\ line flux measured in \jykms. Distances to the galaxies were calculated based on their recessional velocities adopting $H_0=70~\hubble$.
For the galaxies in which we did not detect \hi\ emission, we calculated conservative upper limits to the \hi\ mass, M$_{\rm HI, min}$, by assuming a 5$\sigma$ detection limit over a width of 100~\kms, where the data are smoothed to 30~\kms. 

The GBT beam at 1420~GHz has an approximate FWMH of $10'$. Based on the optical images of the galaxies in our sample, we can safely assume that all \hi\ emission is included in the beam. The physical scale of the beam ranges from 60 kpc for the nearest galaxy to 600 kpc for the galaxy at the largest distance.

The detected \hi\ emission is in most cases nicely centered around the optical recessional velocity of the galaxies, except for the case of CGCG~017-079. The \hi\ emission line velocity is $4263\pm2$~\kms, whereas the published optical redshifts range from 4395 to 4456~\kms, with typical uncertainties of a few tens of \kms. It is therefore suggestive that the \hi\ emission is offset from the optical galaxy. Subsequent higher resolution JVLA spatial mapping of the \hi\ emission in this galaxy has shown that the \hi\ emission is actually confused by the \hi\ emission from the nearby face-on spiral galaxy NGC~5327 to the south of CGCG~017-079. These observations will be presented in a forthcoming paper, but we have used the JVLA data to separate the \hi\ emission of CGCG~017-079 and NGC~5327. The data in Table~\ref{obstable.tab} for this galaxy are therefore not derived from the GBT spectra, but from the JVLA data. In Figure~\ref{spectra.fig} the spectrum of CGCG~017-079 is indicated by a blue line, which shows that the peak of the \hi\ spectrum coincides with the optical redshift. 

The shapes of the \hi\ profiles look very typical for those of spiral galaxies. Two of our detections display a pronounced double-horned profile (2dFGRS-S833Z022 and ESO471-G032), suggesting the presence of a rotating gas disk. The other four detections are consistent with being either flat-topped or mildly double-peaked. These profile shapes suggest that the \hi\ gas is settled in ordered rotation. We will come back to this point in section~\ref{widths.sec}. Interestingly, the optical images in Fig.~\ref{spectra.fig} hint to the fact that the \hi-detected galaxies are preferentially of late morphological type. We cannot draw any firm conclusions as to the relation between environment and \hi\ detection rate: two out of five field E+A galaxies are detected and four out of six E+A galaxies from higher density regions have detectable \hi. Clearly, a larger sample is required to address this question.

\subsection{Previous \hi\ measurements in E+A galaxies}
The first detection of the \hi\ 21-cm emission line in an E+A galaxy was made by \citet{Chang2001a}, who performed a VLA spectral line survey of five E+A galaxies selected from the \citet{Zabludoff1996a} sample. Due to the nature of their selection, these galaxies are at relatively large redshifts of between $z=0.08$ and $0.12$, which makes deep \hi\ line emission searches very time-consuming. One E+A galaxy was detected, with an \hi\ mass of $7.7\times 10^9~\msol$. For the four other galaxies, upper limits to \mhi\ of between $\sim2$ and $\sim5\times 10^9~\msol$ were obtained. The \hi\ detected galaxy, named "EA 1", shows a disturbed gas morphology with tidal tails, suggestive of a recent merger. \citet{Buyle2008a} obtained higher spatial  resolution \hi\ maps and concluded that the galaxy is interacting with its optical companion. Interestingly, the positions of both EA~1 and its companion correspond to minima in the  \hi\ column density, indicating that in the immediate surroundings of the galaxies the gas has been consumed in the last episode of star formation, and no gas is available for ongoing star formation.  

A second search for \hi\ emission in E+A galaxies is the one from \citet{Buyle2006a}, who selected the three nearest galaxies from the SDSS sample of \citet{Goto2003a} and the three nearest galaxies from \citet{Zabludoff1996a}, to perform single dish 21-cm observations with the Parkes and Arecibo telescopes. These galaxies are at redshifts between 0.05 and 0.1. Four of the six galaxies were detected, and were found to have \hi\ masses between $1$ and $7\times 10^9~\msol$.

\citet{Helmboldt2007a} carried out an \hi\ search for E+A galaxies from the SDSS, but his selection criteria were much less stringent than those of most other E+A samples. For this sample it was required that EW(H$\delta$)$>2$\AA\ in absorption, implying the inclusion of many galaxies with much weaker Balmer absorption lines than the galaxies in the samples of, e.g., \citet{Goto2007a}, \citet{Blake2004a}, and \citet{Zabludoff1996a}. It is therefore doubtful whether these galaxies can be regarded as genuine E+A galaxies and we decided to not include this sample in the current analysis.

The  number of all \hi\ emission line detections resulting from targeted 21-cm surveys of bona-fide E+A galaxies, including the surveys of  \citet{Chang2001a}, \citet{Buyle2006a}, and the present GBT survey, now totals 11. Table~\ref{allhi.tab} summarizes the properties of these 11 galaxies. Altogether, 22 galaxies have been surveyed in these three surveys, implying a detection rate of 50 per cent, although it should be noted that the detection limits vary between the different surveys.

\begin{table*}
\centering
\caption{Summary of all published \hi\ detections in E+A galaxies}
\begin{tabular}{l c c c c c c c c c}
\hline
 galaxy name &  source &  $M_B$ & $M_K$ & \mhi\                 &  $V_{\rm hel}$ &  $W_{20}$ & $S_{\rm int}$ & EW(H$\delta$) & environment\\
                      &              &      (mag)  & (mag)     &  ($10^8$ \msol)  &  (\kms)             &  (\kms)         & (\jykms)           & (\AA) & \\
                    (1) & (2) & (3) & (4) & (5) & (6) & (7) & (8) & (9) & (10)\\  

\hline\hline
         SDSS J210258.87+103300.6 & 1 & -21.7 & -25.8 & 65 & 27821 & 440 & 0.18 &  5.1 & 	Field\\ 
         SDSS J230743.41+152558.4 & 1 & -20.4 & -23.5 & 9.0 & 20894 & 240 & 0.04 &  6.8 & 	Cluster\\ 
         SDSS J233453.20+145048.7 & 1 & -20.2 & -23.8 & 27 & 19388 & 380 & 0.15 & 5.2 & 	Field\\ 
            LCRS B002018.8-415015 & 1 & -18.9 & -23.9 & 23 & 17941 & 660 & 0.15 &  6.0 & 	Field \\ 
                             EA 1 & 2 & -20.0 & -24.2 & 77 & 22380 & 191 & 0.26 &  9.0 & 			Field \\ 
                     CGCG 017-079 & 3 & -18.4 & -21.0 & 4.1 &  4437 & 188 & 0.44 & 15.6 & 		Poor cluster/group \\ 
                  2dFGRS S833Z022 & 3 & -16.1 & ... & 15 &  2194 & 184 & 6.35 &  9.2 & 			Field \\ 
                      ESO405-G015 & 3 & -20.4 & ... & 55 &  9938 & 400 & 1.16 &  7.4 & 			Cluster \\ 
                      ESO471-G032 & 3 & -19.4 & -24.4 & 60 &  8637 & 394 & 1.66 &  6.0 & 		Cluster \\ 
                      ESO534-G001 & 3 & -15.8 & ... & 9.6 &  1407 & 137 & 10.10 &  9.4 & 			Field \\ 
         SDSS J162702.64+432833.9 & 3 & -20.8 & -24.9 & 44 & 13796 & 496 & 0.48 &  5.7 & 	Cluster/infall region \\ 
\hline
\end{tabular}
\label{allhi.tab}
\\{\small {\em Notes:} 
column (1): galaxy name;
column (2): source of the \hi\ detection: 1: \citet{Buyle2006a}, 2: \citet{Chang2001a}, 3: this paper;
column (3): $B$-band absolute magnitude;
column (4): $K$-band absolute magnitude taken from the 2MASS Point Source Catalogue where available \citep{Skrutskie2006a};
column (5): measured total \hi\ mass;
column (6): heliocentric recessional velocity measured in \hi;
column (7): width of 21-cm profiles measured at 20 per cent of peak intensity;
column (8): integrated \hi\ 21-cm flux;
column (9): equivalent width of the H$\delta$ line in absorption;
column (10): galaxy environment.}
\end{table*}%

\subsection{The gas-richness of E+A galaxies}
\begin{figure}
\begin{center}
\includegraphics[width=7.5cm,trim=0.cm 0.0cm 0.cm 0.cm]{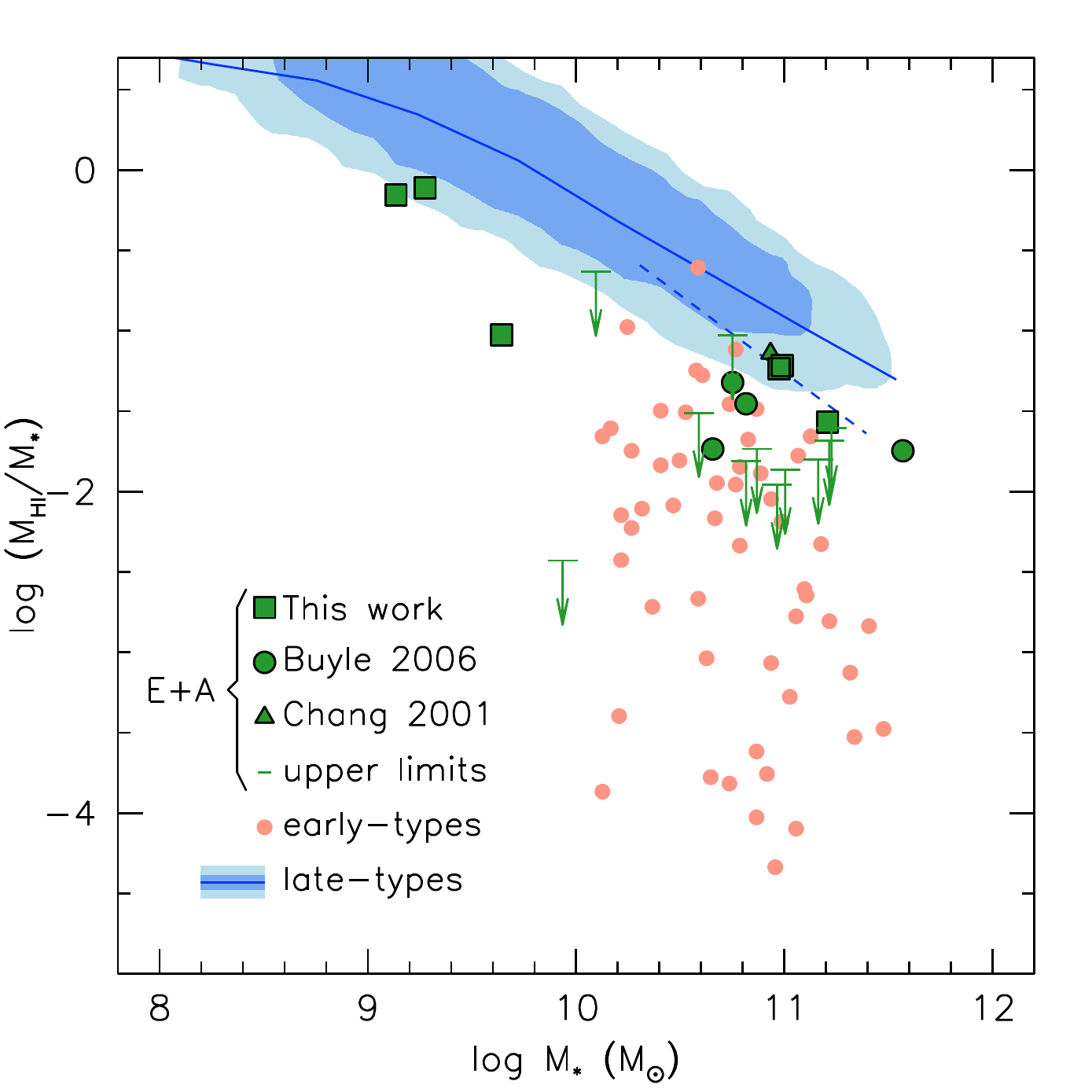}
\caption{Gas fractions $\mhi/M_*$ plotted as a function of stellar mass $M_*$. E+A galaxies are shown as green points. The early-type galaxies from \citet{Serra2012a} are indicated by red circles, and the mean value for field-galaxies from \citet{Huang2012a} is represented by a blue solid line, where the light and dark areas delineate the regions containing 80 and 40 per cent of the data points. The blue dashed line shows the relation from \citet{Catinella2012a} for a complete sample of stellar mass-selected galaxies.
 \label{gas_star.fig}}
\end{center}
\end{figure}

It is now firmly established that a significant fraction of E+A galaxies contains a detectable amount of \hi\ gas. From our sample, in combination with the previous samples studied by \citet{Buyle2006a} and \citet{Chang2001a}, we can conclude that at least half of the nearby E+A galaxies have an \hi\ gas reservoir. The exact fraction is difficult to ascertain because clearly this number must be a function of the depth of \hi\ 21-cm survey. Here, we examine how the gas richness of E+A galaxies compares to that of blue cloud and red sequence galaxies in the local Universe. Previous studies have made this comparison based on the ratio of \hi\ mass to $B$-band luminosity \citep{Buyle2006a} and concluded that E+A galaxies detected in \hi\ 21-cm are nearly as gas rich as spiral galaxies with comparable luminosities. However, using gas fractions calculated from $B$-band luminosities may bias the results toward E+A appearing too gas-rich because E+A galaxies have on average redder colours that typical spiral galaxies. A fairer comparison would be to use gas fractions based on stellar mass estimates such that the true baryonic components in cold gas and stars are compared. Hence, we define gas richness as the \hi\ mass divided by the stellar mass. In Figure~\ref{gas_star.fig} all E+A galaxies listed in Table~\ref{allhi.tab} are shown as green points, as well as the upper limits obtained from the various 21-cm surveys. 

For the E+A galaxies, we estimate stellar masses from 2MASS $K$-band luminosities and assuming a mass composition of 10 per cent young (1 Gyr) and 90 per cent old (10 Gyr) stars. The mass-to-light ratios are taken from \citet{Maraston2005a} for a Salpeter IMF and solar metallicity. For five galaxies $K$-band data are not available and we use $B$-band luminosities instead. Changing the young to old stellar mass ratio to 30--70 per cent would reduce the mass estimates by a factor of $\sim$2. Using a Kroupa IMF would also reduce the mass estimates by a factor of $\sim$1.4.

The gas-richness of the local field galaxy population has been studied in a statistical sense for many years, initially summarized by \citet{Roberts1994a}. With the advent of large-scale optical and 21-cm surveys, the statistics have improved significantly over the last few years, resulting in the studies of, e.g.,  \citet{Catinella2012a}, who concentrate on Arecibo observations of high-mass galaxies ($M_*>10^{10} \msol$), \citet{West2010a}, based on the HIPASS survey, \citet{Bothwell2009a}, using literature values for a very local galaxy sample, and \citet{Huang2012a} based on the ALFALFA survey. Although these surveys do not specifically target late-type galaxies, the correlations between gas richness and stellar mass are dominated by this type. It is well established that for late-type galaxies the gas-richness is a strong negative function of galaxy mass. From \citet{Huang2012a} it can be seen that the gas-richness $\mhi/M_*$ varies from $\sim$10 for galaxies with $M_*=10^8\msol$ to ~0.1 for  $M_*=10^{11}\msol$. In Figure~\ref{gas_star.fig} we show as a blue line their parametrized mean gas richness, with dark and light coloured areas indicating the regions that include approximately 80 and 40 per cent of the data points. The \citet{Huang2012a} stellar masses are based on a \citet{Chabrier2003a} initial mass function (IMF). To convert to a Salpeter IMF, we multiplied the stellar masses by a factor 1.4, as per \citet{Longhetti2009a}.

Also shown as a dashed line are the results from \citet{Catinella2012a} for galaxies with stellar masses $>10^{10}~\msol$. The difference between this curve and the one based on the \citet{Huang2012a} results is that the latter is essentially based on an \hi\ selected galaxy sample, whereas \citet{Catinella2012a} obtained \hi\ measurements of a complete sample of stellar mass selected galaxies. The blue area in Figure~\ref{gas_star.fig} might therefore be biased toward higher gas fractions.

To compare with early-type galaxies, we use the recent compilation of \citet{Serra2012a}, who studied the \hi\ distribution in a sample of 166 nearby early-type galaxies drawn from the ATLAS$^{\rm 3D}$ survey catalogue. The ATLAS$^{\rm 3D}$ sample is a volume (D\,$< 42$\,Mpc) and magnitude limited ($M_K < 21.5$\,mag) sample in the northern sky. In this sample, 40 per cent of the galaxies outside the Virgo cluster and 10 per cent within it have detectable \hi\ gas. The overall gas richness of the gas-bearing early types is lower than that of spirals, but there is a large spread, with some early-type type galaxies having gas fractions comparable to spiral galaxies. In Figure~\ref{gas_star.fig} we add the points from the \citet{Serra2012a} sample as red closed points. For the conversion from $K$-band luminosity to stellar mass, we adopt an overall mass-to-light ratio of $M/L_K=1.4$.

We conclude that gas-bearing E+A galaxies have cold gas fractions similar to the most gas-rich early type galaxies and the most gas-poor late-type galaxies. The combined sample of E+As with \hi\ detections spans a very large range in stellar mass and, interestingly,  the offset from the gas richness--stellar mass relation for late-type galaxies appears to be consistent over the whole range.   This observation is consistent with the idea the E+A galaxies are in a transitional phase between gas-rich actively star forming galaxies and quiescent early-type galaxies. The starburst that must have happened in the last ~Gyr has consumed only a fraction of the total available \hi\ reservoir.

\begin{figure}
\begin{center}
\includegraphics[width=7.5cm,trim=0.cm 0.0cm 0.cm 0.cm]{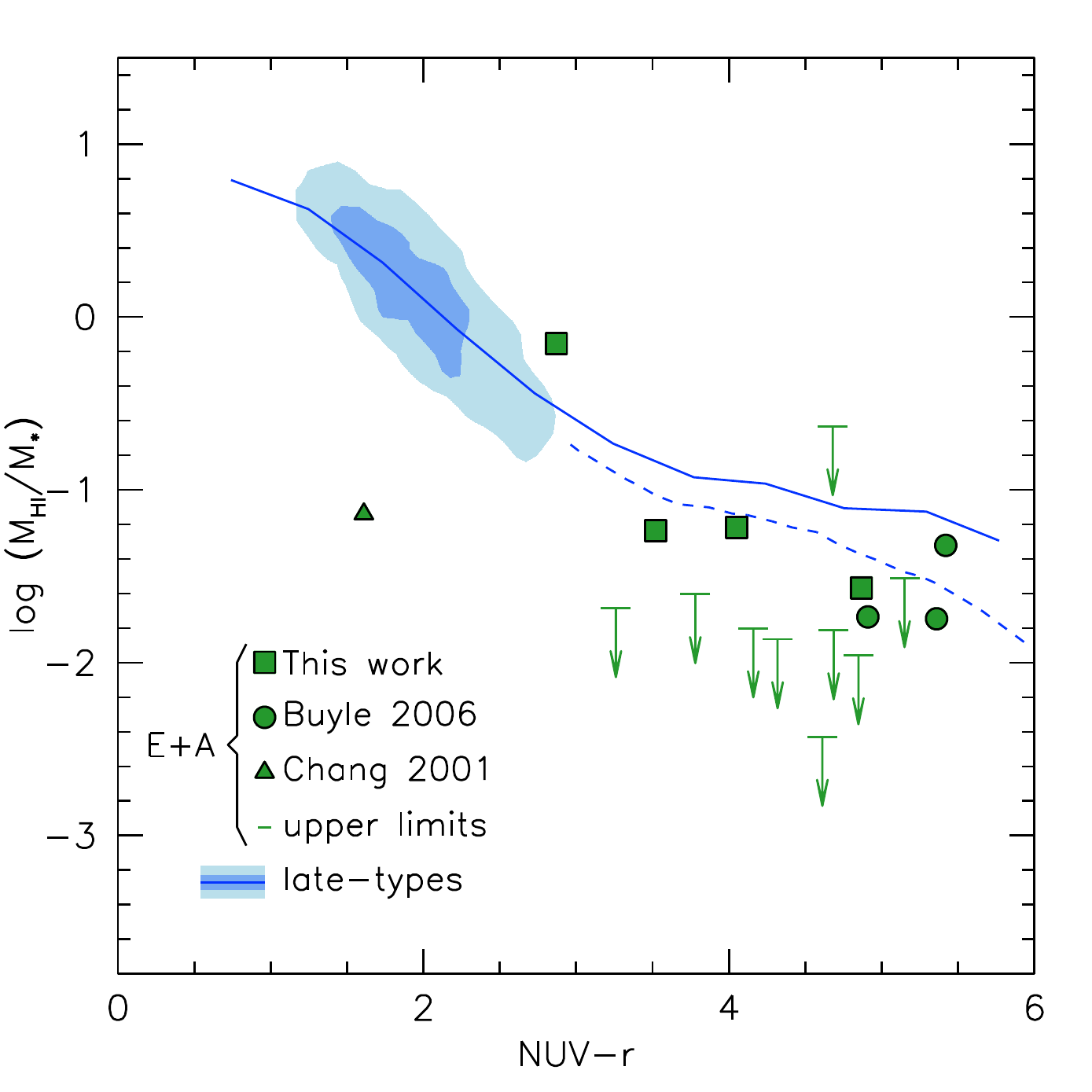}
\caption{Gas fractions $\mhi/M_*$ plotted as a function of  NUV-$r$ colour. E+A galaxies are shown as green points. The mean value for field-galaxies from \citet{Huang2012a} is represented by a blue solid line, where the light and dark areas lines delineate the regions containing 50 and 30 per cent of the data points. The dashed line corresponds to the high stellar mass ($>10^{10}\msol$) galaxies from the \citet{Catinella2012a} sample.
 \label{nuvr.fig}}
\end{center}
\end{figure}

There have been several studies recently with the aim of finding reliable estimators of gas-richness based on optical, infrared and UV-measurements. The aim of these studies is to define  'photometric gas fractions' that can be used as a substitute for real cold gas measurement for large samples of galaxies. For example, \citet{Kannappan2004a} argued that the $u-K$ colour of galaxies is a good indicator of their atomic gas fraction. Similar conclusions were drawn by \citet{Huang2012a} and \citet{Zhang2009a}, using variations of the colour indicator, particularly by using NUV$-r$. The explanation for the tight correlation between gas fraction and colour is that the $r$ or $K$-band flux is a measurement of the total stellar mass, while the $u$ or NUV flux acts as a tracer of young stars, and through the Kennicutt-Schmidt relation is an indicator of the surface density of cold gas. 

Here, we test how well we can estimate the cold gas fraction of E+A galaxies based on the NUV$-r$ colour. If the \hi\ gas in E+A galaxies is not related to active star formation, then the expectation may be that E+As are off the photometric gas fraction relation defined for normal gas-rich galaxies. We take NUV magnitudes from the GALEX GR6 Data Release\footnote{http://galex.stsci.edu/GR6/} and $r$-band measurements from NED\footnote{http://ned.ipac.caltech.edu/}. In total, we obtain joint stellar mass and NUV$-r$ measurements of eight of the eleven galaxies of the combined E+A sample with \hi\ detections. 

Figure~\ref{nuvr.fig} shows the relation for our combined sample of E+As, together with the relation for gas-rich galaxies from \citet{Huang2012a}. The dashed line is again the relation from \citet{Catinella2012a}. We find that nearly all E+A galaxies with {\em detected} \hi\ are essentially on the gas fraction -- (NUV$-r$) relation, apart from galaxy EA~1 from \citet{Chang2001a}. As discussed before, this latter galaxy is actually a close pair undergoing a merger \citep{Buyle2008a}. For all other \hi\ detected galaxies in the sample, the atomic gas fraction can be estimated from their NUV-$r$ colour with an accuracy of $~0.3$ dex. What we cannot establish from our present sample is whether the \hi-undetected E+A galaxies populate the  region directly underneath the relation for spiral galaxies and that the \hi-detected galaxies form an upper envelope, or, alternatively, that approximately half the population of E+A galaxies follow the normal gas fraction -- (NUV$-r$) relation and the other half contains no cold gas at all. Deeper \hi\ surveys of a complete sample of E+A galaxies are required to answer this question.

\subsection{Velocity widths}\label{widths.sec}
\begin{figure}
\begin{center}
\includegraphics[width=7.5cm,trim=0.cm 0.0cm 0.cm 0.cm]{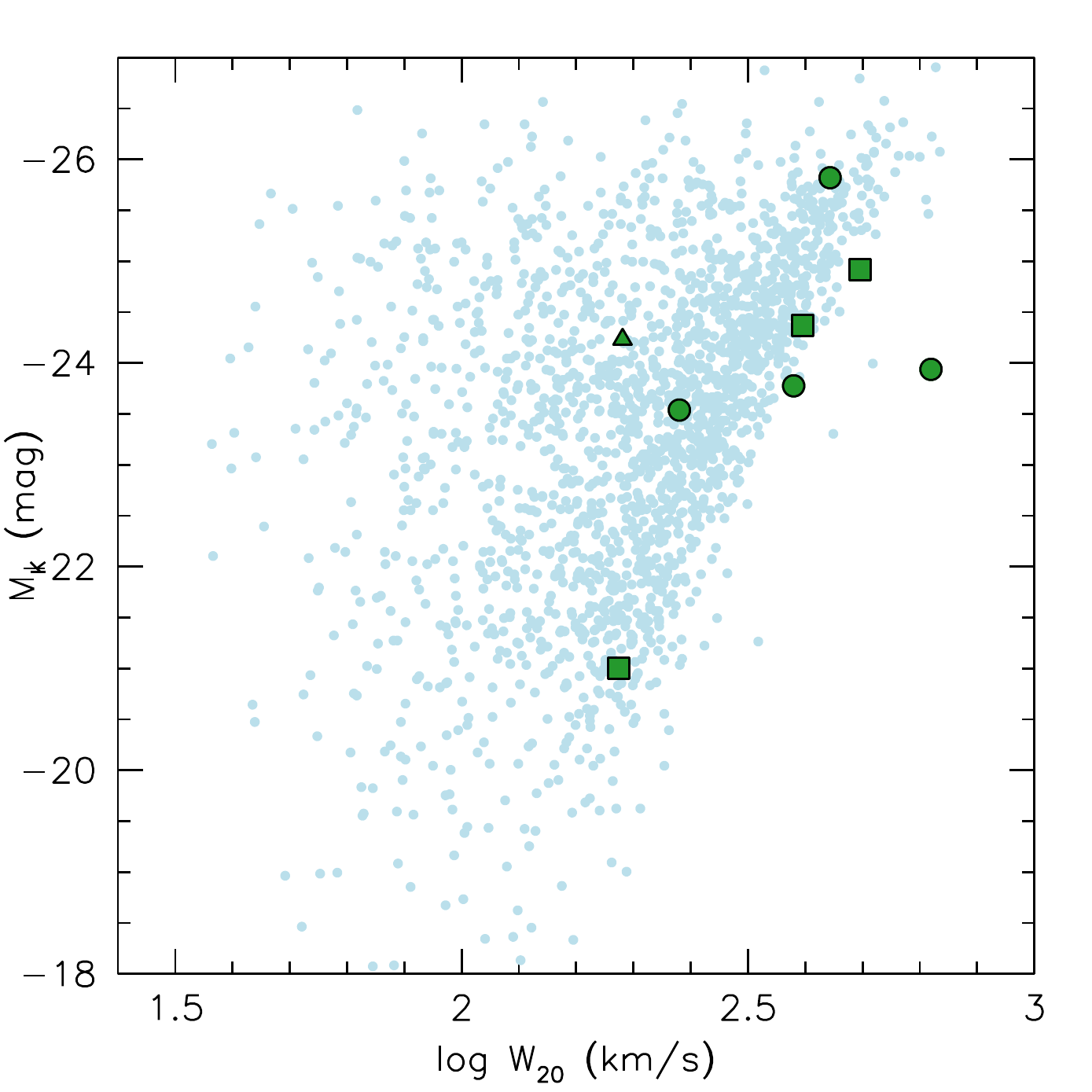}
\caption{$K$-band absolute magnitude plotted against  \hi\ velocity widths uncorrected for inclination. Small faint dots are \hi\ selected galaxies from \citet{Meyer2008a} and large symbols are E+A galaxies. Symbols are equal to those in Figures~\ref{gas_star.fig} and \ref{nuvr.fig}.
 \label{TF.fig}}
\end{center}
\end{figure}

Finally, we discuss the velocity widths of the gas detected in E+A galaxies. In Table~\ref{allhi.tab} the widths of the \hi\ profiles measured at 20 per cent of the peak flux are listed. Ideally, to make a fair comparison with the rotational velocities of spiral galaxies, an inclination correction would have to be applied. However, given the mixed morphologies of our sample, reliable inclination angles are impossible to measure. Instead, we choose to compare with measured uncorrected velocity widths from the literature. \citet{Meyer2008a} present a $K$-band-based Tully-Fisher analysis of \hi-selected galaxies. In Figure~\ref{TF.fig} we reproduce their measurements, but plot measured profiles widths uncorrected for inclination on the horizontal axis. The sample of E+A galaxies is overplotted, with symbols equal to those in Figures~\ref{gas_star.fig} and \ref{nuvr.fig}. The \citet{Meyer2008a} galaxies define a sharp envelope which is determined by edge-on galaxies for which the full rotational velocity is measured. The E+A galaxies follow a very similar distribution, suggesting that the \hi\ gas is fully sampling the galaxies' potential wells. In other words, this plot is suggesting that in most E+A galaxies the cold gas is settled in a disk, and not in clouds that are the result of interactions. The only clear exception is LCRS B002018.8-415015, which with a velocity width of 660~\kms\ has a cold gas velocity spread a factor two larger than allowed by the Tully-Fisher relation. It should be noted that this galaxy has a very likely companion at a projected distance of $\sim$18~kpc, so the  single dish \hi\ detection could be confused.

Although Figure~\ref{TF.fig} lends support to the idea that the cold gas in many E+A galaxies is settled in ordered rotation, this is clearly not the case for all E+As. As discussed in section \ref{results.sec}, many of the 21-cm profiles are symmetric, but even the profile of EA1 presented in \citet{Buyle2008a} is symmetric, although this gas is clearly disturbed.

\section{Conclusions}
We have performed a new Green Bank Telescope 21-cm line emission survey for \hi\ gas in post-starburst, or E+A, galaxies.  Our sample galaxies are at very low distances ($z<0.05$), which allows us to achieve sensitive cold gas mass measurements. Our conclusions are the following:
\begin{enumerate}
\item \hi\ emission is detected in six out of eleven galaxies surveyed. Combined with earlier studies, the total number of E+A galaxies with measured cold gas components is now eleven. Approximately half of the nearby E+A population has a detectable cold gas component ($>$10 per cent of the equivalent gas mass of spiral galaxies).
\item Gas-bearing E+A galaxies have cold gas fractions similar to the most gas-rich early type galaxies and the most gas-poor late-type galaxies. Over the very large range in stellar mass spanned by the combined sample of E+As with \hi\ detections, the offset from the gas richness--stellar mass relation for late-type galaxies appears to be constant.
\item Gas-bearing E+A galaxies follow the "photometric gas fraction" estimator based on the NUV-$r$ colour, defined for normal field galaxies. However, this relation only holds for the $\sim$50 per cent of E+As with detectable cold gas.
\item The 21-cm profile shapes of the E+A galaxies are suggestive of gas settled in an ordered disk. The relation between $K$-band luminosity and observed \hi\ velocity width is consistent with this idea.
\end{enumerate}
These conclusions are consistent with the idea that E+A galaxies form a transitional population between star-forming, disk-dominated galaxies and spheroidal quiescent, non-star forming galaxies. However, a number of questions remain. Prominent among these is the question of why the star formation has ceased in the last $\sim$1 Gyr if large amounts of gas are still available? If the cold gas resides in an ordered disk, then why is this gas stable against forming stars? To answer these questions, the distribution and kinematics of the \hi\ and the molecular gas needs to be studied in more detail.

\section*{Acknowledgments}
We thank the referee Pieter Buyle for his comments that improved the paper. WJC acknowledges the financial support of the Australian Research Council and the award of an Australian Professorial Fellowship throughout the course of this work. The results presented in this paper  came from  program GBT09C-005. We thank the GBT staff for their support. The National Radio Astronomy Observatory is a facility of the National Science Foundation operated under cooperative agreement by Associated Universities, Inc. 

\small
\bibliographystyle{mnras}
\bibliography{/Users/mzwaan/REFERENCES/zwaanreferences}

\begin{thebibliography}{}

\bibitem[\protect\citeauthoryear{{Bell} et~al.}{{Bell}
  et~al.}{2004}]{Bell2004a}
{Bell} E.~F. et~al., 2004, \apj, 608, 752

\bibitem[\protect\citeauthoryear{{Blake} et~al.}{{Blake}
  et~al.}{2004}]{Blake2004a}
{Blake} C. et~al., 2004, \mnras, 355, 713

\bibitem[\protect\citeauthoryear{{Blanton} \& {Moustakas}}{{Blanton} \&
  {Moustakas}}{2009}]{Blanton2009a}
{Blanton} M.~R.,  {Moustakas} J., 2009, \araa, 47, 159

\bibitem[\protect\citeauthoryear{{Bothwell}, {Kennicutt} \& {Lee}}{{Bothwell}
  et~al.}{2009}]{Bothwell2009a}
{Bothwell} M.~S., {Kennicutt} R.~C.,  {Lee} J.~C., 2009, \mnras, 400, 154

\bibitem[\protect\citeauthoryear{{Buyle}, {De Rijcke} \& {Dejonghe}}{{Buyle}
  et~al.}{2008}]{Buyle2008a}
{Buyle} P., {De Rijcke} S.,  {Dejonghe} H., 2008, \apjl, 684, L17

\bibitem[\protect\citeauthoryear{{Buyle} et~al.}{{Buyle}
  et~al.}{2006}]{Buyle2006a}
{Buyle} P., {Michielsen} D., {De Rijcke} S., {Pisano} D.~J., {Dejonghe} H.,
  {Freeman} K., 2006, \apj, 649, 163

\bibitem[\protect\citeauthoryear{{Caldwell} et~al.}{{Caldwell}
  et~al.}{1996}]{Caldwell1996a}
{Caldwell} N., {Rose} J.~A., {Franx} M.,  {Leonardi} A.~J., 1996, \aj, 111, 78

\bibitem[\protect\citeauthoryear{{Catinella} et~al.}{{Catinella}
  et~al.}{2012}]{Catinella2012a}
{Catinella} B. et~al., 2012, \aap, 544, A65

\bibitem[\protect\citeauthoryear{{Chabrier}}{{Chabrier}}{2003}]{Chabrier2003a}
{Chabrier} G., 2003, PASP, 115, 763

\bibitem[\protect\citeauthoryear{{Chang} et~al.}{{Chang}
  et~al.}{2001}]{Chang2001a}
{Chang} T.-C., {van Gorkom} J.~H., {Zabludoff} A.~I., {Zaritsky} D.,  {Mihos}
  J.~C., 2001, \aj, 121, 1965

\bibitem[\protect\citeauthoryear{{Couch} et~al.}{{Couch}
  et~al.}{1998}]{Couch1998a}
{Couch} W.~J., {Barger} A.~J., {Smail} I., {Ellis} R.~S.,  {Sharples} R.~M.,
  1998, \apj, 497, 188

\bibitem[\protect\citeauthoryear{{Couch} \& {Sharples}}{{Couch} \&
  {Sharples}}{1987}]{Couch1987a}
{Couch} W.~J.,  {Sharples} R.~M., 1987, \mnras, 229, 423

\bibitem[\protect\citeauthoryear{{Dressler} \& {Gunn}}{{Dressler} \&
  {Gunn}}{1983}]{Dressler1983a}
{Dressler} A.,  {Gunn} J.~E., 1983, \apj, 270, 7

\bibitem[\protect\citeauthoryear{{Dressler} et~al.}{{Dressler}
  et~al.}{1999}]{Dressler1999a}
{Dressler} A., {Smail} I., {Poggianti} B.~M., {Butcher} H., {Couch} W.~J.,
  {Ellis} R.~S.,  {Oemler} A., Jr., 1999, \apjs, 122, 51

\bibitem[\protect\citeauthoryear{{Emsellem} et~al.}{{Emsellem}
  et~al.}{2011}]{Emsellem2011a}
{Emsellem} E. et~al., 2011, \mnras, 414, 888

\bibitem[\protect\citeauthoryear{{Faber} et~al.}{{Faber}
  et~al.}{2007}]{Faber2007a}
{Faber} S.~M. et~al., 2007, \apj, 665, 265

\bibitem[\protect\citeauthoryear{{Garcia}}{{Garcia}}{1993}]{Garcia1993a}
{Garcia} A.~M., 1993, \aaps, 100, 47

\bibitem[\protect\citeauthoryear{{Goto}}{{Goto}}{2007}]{Goto2007a}
{Goto} T., 2007, \mnras, 381, 187

\bibitem[\protect\citeauthoryear{{Goto} et~al.}{{Goto}
  et~al.}{2003}]{Goto2003a}
{Goto} T. et~al., 2003, \pasj, 55, 771

\bibitem[\protect\citeauthoryear{{Helmboldt}}{{Helmboldt}}{2007}]{Helmboldt2007a}
{Helmboldt} J.~F., 2007, \mnras, 379, 1227

\bibitem[\protect\citeauthoryear{{Huang} et~al.}{{Huang}
  et~al.}{2012}]{Huang2012a}
{Huang} S., {Haynes} M.~P., {Giovanelli} R.,  {Brinchmann} J., 2012, \apj, 756,
  113

\bibitem[\protect\citeauthoryear{{Jarosik} et~al.}{{Jarosik}
  et~al.}{2011}]{Jarosik2011a}
{Jarosik} N. et~al., 2011, \apjs, 192, 14

\bibitem[\protect\citeauthoryear{{Kannappan}}{{Kannappan}}{2004}]{Kannappan2004a}
{Kannappan} S.~J., 2004, \apjl, 611, L89

\bibitem[\protect\citeauthoryear{{Lauberts}}{{Lauberts}}{1982}]{Lauberts1982a}
{Lauberts} A., 1982, {ESO/Uppsala survey of the ESO(B) atlas}

\bibitem[\protect\citeauthoryear{{Longhetti} \& {Saracco}}{{Longhetti} \&
  {Saracco}}{2009}]{Longhetti2009a}
{Longhetti} M.,  {Saracco} P., 2009, \mnras, 394, 774

\bibitem[\protect\citeauthoryear{{Maraston}}{{Maraston}}{2005}]{Maraston2005a}
{Maraston} C., 2005, \mnras, 362, 799

\bibitem[\protect\citeauthoryear{Meyer et~al.}{Meyer et~al.}{2008}]{Meyer2008a}
Meyer M.~J., Zwaan M.~A., Webster R.~L., Schneider S.,  Staveley-Smith L.,
  2008, \mnras, 391, 1712

\bibitem[\protect\citeauthoryear{{Norton} et~al.}{{Norton}
  et~al.}{2001}]{Norton2001a}
{Norton} S.~A., {Gebhardt} K., {Zabludoff} A.~I.,  {Zaritsky} D., 2001, \apj,
  557, 150

\bibitem[\protect\citeauthoryear{{Poggianti} et~al.}{{Poggianti}
  et~al.}{1999}]{Poggianti1999a}
{Poggianti} B.~M., {Smail} I., {Dressler} A., {Couch} W.~J., {Barger} A.~J.,
  {Butcher} H., {Ellis} R.~S.,  {Oemler} A., Jr., 1999, \apj, 518, 576

\bibitem[\protect\citeauthoryear{{Pracy.} et~al.}{{Pracy.}
  et~al.}{2013}]{Pracy2013a}
{Pracy.} M.~B., Croom S., {Sadler} E.~M., {Couch} W.~J., {Kuntschner} H.,
  {Owers} M.~S.,  Zwaan M.~A., 2013, MNRAS, submitted

\bibitem[\protect\citeauthoryear{{Pracy} et~al.}{{Pracy}
  et~al.}{2009}]{Pracy2009a}
{Pracy} M.~B., {Kuntschner} H., {Couch} W.~J., {Blake} C., {Bekki} K.,
  {Briggs} F., 2009, \mnras, 396, 1349

\bibitem[\protect\citeauthoryear{{Pracy} et~al.}{{Pracy}
  et~al.}{2012}]{Pracy2012a}
{Pracy} M.~B., {Owers} M.~S., {Couch} W.~J., {Kuntschner} H., {Bekki} K.,
  {Briggs} F., {Lah} P.,  {Zwaan} M., 2012, \mnras, 420, 2232

\bibitem[\protect\citeauthoryear{{Rines} et~al.}{{Rines}
  et~al.}{2002}]{Rines2002a}
{Rines} K., {Geller} M.~J., {Diaferio} A., {Mahdavi} A., {Mohr} J.~J.,
  {Wegner} G., 2002, \aj, 124, 1266

\bibitem[\protect\citeauthoryear{{Roberts} \& {Haynes}}{{Roberts} \&
  {Haynes}}{1994}]{Roberts1994a}
{Roberts} M.~S.,  {Haynes} M.~P., 1994, \araa, 32, 115

\bibitem[\protect\citeauthoryear{{Serra} et~al.}{{Serra}
  et~al.}{2012}]{Serra2012a}
{Serra} P. et~al., 2012, \mnras, 422, 1835

\bibitem[\protect\citeauthoryear{{Skrutskie} et~al.}{{Skrutskie}
  et~al.}{2006}]{Skrutskie2006a}
{Skrutskie} M.~F. et~al., 2006, \aj, 131, 1163

\bibitem[\protect\citeauthoryear{{Tinker}, {Wechsler} \& {Zheng}}{{Tinker}
  et~al.}{2010}]{Tinker2010a}
{Tinker} J.~L., {Wechsler} R.~H.,  {Zheng} Z., 2010, \apj, 709, 67

\bibitem[\protect\citeauthoryear{{West} et~al.}{{West}
  et~al.}{2010}]{West2010a}
{West} A.~A., {Garcia-Appadoo} D.~A., {Dalcanton} J.~J., {Disney} M.~J.,
  {Rockosi} C.~M., {Ivezi{\'c}} {\v Z}., {Bentz} M.~C.,  {Brinkmann} J., 2010,
  \aj, 139, 315

\bibitem[\protect\citeauthoryear{{White} et~al.}{{White}
  et~al.}{1999}]{White1999a}
{White} R.~A., {Bliton} M., {Bhavsar} S.~P., {Bornmann} P., {Burns} J.~O.,
  {Ledlow} M.~J.,  {Loken} C., 1999, \aj, 118, 2014

\bibitem[\protect\citeauthoryear{{Williams} et~al.}{{Williams}
  et~al.}{2009}]{Williams2009a}
{Williams} R.~J., {Quadri} R.~F., {Franx} M., {van Dokkum} P.,  {Labb{\'e}} I.,
  2009, \apj, 691, 1879

\bibitem[\protect\citeauthoryear{{Yang} et~al.}{{Yang}
  et~al.}{2008}]{Yang2008a}
{Yang} Y., {Zabludoff} A.~I., {Zaritsky} D.,  {Mihos} J.~C., 2008, \apj, 688,
  945

\bibitem[\protect\citeauthoryear{{Zabludoff} et~al.}{{Zabludoff}
  et~al.}{1996}]{Zabludoff1996a}
{Zabludoff} A.~I., {Zaritsky} D., {Lin} H., {Tucker} D., {Hashimoto} Y.,
  {Shectman} S.~A., {Oemler} A.,  {Kirshner} R.~P., 1996, \apj, 466, 104

\bibitem[\protect\citeauthoryear{{Zhang} et~al.}{{Zhang}
  et~al.}{2009}]{Zhang2009a}
{Zhang} W., {Li} C., {Kauffmann} G., {Zou} H., {Catinella} B., {Shen} S., {Guo}
  Q.,  {Chang} R., 2009, \mnras, 397, 1243

\bibitem[\protect\citeauthoryear{{Zwaan}}{{Zwaan}}{2000}]{Zwaan2000c}
{Zwaan} M.~A., 2000, Ph.D.~Thesis

\end{thebibliography}

\label{lastpage}

\end{document}